\documentclass[conference]{IEEEtran}

\IEEEoverridecommandlockouts

\usepackage{cite}
\usepackage{amsmath}
\usepackage{amsmath,amssymb,amsfonts}
\usepackage{algorithmic}
\usepackage{graphicx}
\usepackage{hyphenat}
\usepackage{multicol}
\usepackage{url}
\usepackage{kpfonts}
\usepackage{xcolor}
\def\BibTeX{{\rm B\kern-.05em{\sc i\kern-.025em b}\kern-.08em
    T\kern-.1667em\lower.7ex\hbox{E}\kern-.125emX}}
   
\begin{document}

\title{Investigation on Machine Learning Based Approaches for Estimating the Critical Temperature of Superconductors
}

\author{
    \IEEEauthorblockA{
        Fatin Abrar Shams\textsuperscript{1,*\thanks{Corresponding author. Email: abrarshams@iut-dhaka.edu}}, Rashed Hasan Ratul\textsuperscript{1,*\thanks{Corresponding author. Email: rashedhasan@iut-dhaka.edu}}, Ahnaf Islam Naf\textsuperscript{1},\\
       Syed Shaek Hossain Samir\textsuperscript{1}, Mirza Muntasir Nishat\textsuperscript{1}, Fahim Faisal\textsuperscript{1} and Md. Ashraful Hoque\textsuperscript{1}
    }
    \IEEEauthorblockA{
        \textsuperscript{1}Department of Electrical and Electronic Engineering\\ Islamic University of Technology (IUT), Dhaka, Bangladesh\\
        \{abrarshams, rashedhasan, ahnafislam, shaekhossain, mirzamuntasir, faisaleee, mahoque\}@iut-dhaka.edu
    }
    
}

\maketitle {
}

\begin{abstract}
Superconductors have been among the most fascinating substances, as the fundamental concept of superconductivity as well as the correlation of critical temperature and superconductive materials have been the focus of extensive investigation since their discovery. However, superconductors at normal temperatures have yet to be identified. Additionally, there are still many unknown factors and gaps of understanding regarding this unique phenomenon, particularly the connection between superconductivity and the fundamental criteria to estimate the critical temperature. To bridge the gap, numerous machine learning techniques have been established to estimate critical temperatures as it is extremely challenging to determine. Furthermore, the need for a sophisticated and feasible method for determining the temperature range that goes beyond the scope of the standard empirical formula appears to be strongly emphasized by various machine-learning approaches. This paper uses a stacking machine learning approach to train itself on the complex characteristics of superconductive materials in order to accurately predict critical temperatures. In comparison to other previous accessible research investigations, this model demonstrated a promising performance with an RMSE of 9.68 and an R2 score of 0.922. The findings presented here could be a viable technique to shed new insight on the efficient implementation of the stacking ensemble method with hyperparameter optimization (HPO).
\end{abstract}

\vspace{1\baselineskip}

\begin{IEEEkeywords}

superconductor, critical temperature, machine learning, stacking ensemble method.

\end{IEEEkeywords}

\section{Introduction}
The significance of superconductivity as a unique and intriguing phenomenon is crucial that has been the focus of intensive study for more than a century \cite{b1}. The critical temperature, Tc, is a temperature below which some particular metals, including indium, mercury, lead, tin, and niobium lose their electrical resistance \cite{b2}. Existing analytical models that attempt to estimate the critical temperature (Tc) from known data are insufficiently reliable as they oversimplify a highly nonlinear and challenging issue. The two unique physical characteristics of superconductors, total electrical conductivity and complete diamagnetism, play critical roles in many industrial manufacturing domains, including mechanical devices, underwater localization sensors, superconducting electric motors, noninvasive part testing, instrumentation, and power systems \cite{b3}. These characteristics of superconductors can also be implemented for designing operational energy distribution systems for network infrastructures with high priority VoLTE (Voice over LTE) services \cite{b25}. Additionally, the high sensitivity of superconductive substances to magnetic fields can be implemented to determine the precise position of underwater sensor nodes with the support of optoacoustic signals \cite{b22}\cite{b21}.
Conventional trial-and-error experimentation for finding novel superconductors takes a long time and necessitates extremely high pressure and low temperatures. Density functional theory (DFT)-based computational techniques are typically also expensive and time-consuming \cite{b4}. Considering the number of existing superconductors and the availability of materials databases, machine learning techniques can be utilized to acquire a deeper understanding of the relationship between superconductivity and the material's chemistry as well as the structure \cite{b5}.\\ It is indeed intriguing to observe that Tcs for the majority of newly discovered superconducting materials, which have a higher two-dimensional electron-phonon interaction, do not match Allen and Dynes' formulation \cite{b6}. Therefore, it appears that the machine-learning methodology is a more efficient approach to underlining the specification that exceeds the limits of the traditional empirical techniques to describe the features of superconductors in terms of critical temperature. As a result, this research presents an integrated machine learning approach to forecast the Tc information of superconducting substances, which can speed up the discovery of prospective high-Tc superconductors. 
Moreover, the performance of a few supervised machine learning models and a stacking ensemble model was tested in order to predict the critical temperature of superconductors based on their physical and chemical properties.\\ In this paper, a unique strategy has been implemented that combines the stacking ensemble method and the hyperparameter optimization, which could be an appealing approach in order to predict the Tc of superconductors. The proposed stacking model indicates superior and stable performance under four conditions that involve variable features and optimization status. Whereas, the other ML models indicate inconsistency and instability in terms of their performance metrics at those conditions. So, the novelty of this study can be considered as follows:\\
• Firstly, the performance parameters remain consistent while applying the proposed stacking ensemble method even after employing the feature reduction technique.\\
• Secondly, compared to other ML approaches, the stacking method shows much more promising results in terms of RMSE and R2 scores.

The rest of the paper is organized as follows. The related works and applicable literature review linked to this work are described in Section 2. In Section 3, the stacking ensemble approach and hyperparameter optimization are described. Sections 4 and 5 respectively concentrate on the results and discussion parts. The concluding remarks are included in Section 6.

\section{Literature Review}
Superconductors are materials that possess zero electric resistance and ideal diamagnetism below a threshold temperature (Tc). Predicting the critical temperature of superconductors is of great interest for various applications in materials science, electronics, and energy storage \cite{b7}. Several research papers have explored the incorporation of machine learning algorithms to predict the Tc of known superconductors. 
Stanev et al. developed machine learning models to predict the critical temperatures (Tc) of known superconductors solely based on their chemical compositions \cite{b8}. The study employed separate regression models for cuprate, iron-based, and low critical temperature-based compounds using data extracted from the SuperCon database. However, it should be noted that the models in this study do not consider other factors that may influence superconductivity, such as crystal structure and electronic properties.
García-Nieto et al. proposed a predictive model to estimate the critical temperature of a superconductor using the physico-chemical properties of the material \cite{b9}. The model combined the multivariate adaptive regression splines (MARS) technique with the whale optimization algorithm (WOA), a meta-heuristic optimization algorithm. While the proposed model achieved promising results, it was based on a specific set of input variables, limiting its applicability to other sets of input variables.
Xie et al. addressed the challenging task of predicting the critical temperature (Tc) of new superconductors using machine learning algorithms \cite{b10}. The study introduced a simple analytical expression obtained through machine learning, which outperformed the traditional Allen-Dynes fit. Additionally, the research highlighted the importance of a new descriptor characterizing Fermi-surface properties to enhance Tc prediction accuracy. Agbemade conducted a study that included feature selection employing the Backwards selection approach, model diagnostics, and a detailed analysis of the distribution of Tc \cite{b11}. After tweaking its hyperparameters, the gradient-boosted method outperformed multiple linear regression with an RMSE value of 12.01 and an R2 score of 0.8823 in terms of effectively estimating Tc in superconducting materials. Le et al. suggested a unique method for forecasting the critical temperature of superconducting materials using the Variational Bayesian Neural Network, a generative machine-learning framework. With encouraging outcomes, this strategy used chemical constituents and superconductor formulas to predict Tc \cite{b12}. Using variational inference, the authors of this work were able to approximate the distribution of parameters in latent features for the generative model, leading to an R2 value of nearly 0.94 and a significant decrease in the RMSE value (3.83). Yu et al. presented a strategy for predicting the critical temperature (Tc) of superconductors based on a two-layer feature choice and Optuna-Stacking ensemble learning model \cite{b13}. The model integrated a filtered and packed selection of features, resulting in data-driven superconductor research that is efficient and cost-effective. Applying Optuna, an autonomous hyperparameter tuning approach, to identify the best parameters for the dataset under consideration is a critical and significant advancement in the discipline of ML \cite{b23}. Roter et al. employed machine learning techniques to predict critical temperatures of superconductors solely based on their chemical composition \cite{b14}. The study achieved a strong coefficient of determination (R2 = 0.93) in the prediction model, comparable to other ML techniques. Hamidieh et al. presented a statistical model in their study to predict the critical temperatures (Tc) of superconductors \cite{b15}. The model achieved an RMSE of 9.5, indicating reasonably accurate predictions. The research incorporated both multiple regression and XGBoost models, harnessing the strengths of these techniques in Tc prediction. \\
Machine learning approaches have shown promise in predicting the critical temperatures (Tc) of superconductors based on their chemical compositions. Various models, including regression models, hybrid algorithms, and generative frameworks, have been explored to improve the accuracy of Tc predictions. Almost all of the approaches are technically complex and computationally demanding. However, these studies contribute valuable insights to the field of materials science and pave the way for further advancements in superconductivity research using machine learning techniques. In this paper, we propose a stacking integrated machine learning technique that combines different simplified machine-learning algorithms and strengthens the metamodel's hyperparameters. As a result, the suggested method offers a more thorough and accurate representation of ML model performance in material discovery applications with a simplified strategy. The study demonstrated a high coefficient of determination (R2 score = 0.922) in the forecasting framework comparable to or better than some previous artificial intelligence methods.

\section{Materials and Methods}
\subsection{Dataset Description}
This dataset emphasizes a cutting-edge machine learning technique to extract complex superconductive material properties for critical temperature (Tc) prediction. The information about superconductors was accumulated from the Superconducting Material Database (SuperCon), and is maintained by the Japanese National Institute of Materials Science (NIMS) \cite{b5}. However, this database was not open-accessed at the time of this research. As a result, the same dataset from the UCI Machine Learning Repository was used \cite{b16}. It encompasses 81 features taken from 21,263 superconductors, as well as the critical temperature.

\subsection{Dataset Preprocessing and Feature Extraction }
Based on the fundamental characteristics of the superconductors, features are developed that could be meaningful for predicting Tc. The properties of an element, such as atomic mass, atomic radius, electron affinity, thermal conductivity, fusion heat, and many others, can be used to assess Tc. There are 81 features that have been defined and extracted from each superconductor. In order to pre-process the dataset, at first, the \textit{min max} scaling method has been implemented for all 81 features in order to normalize the data. 
Secondly, one feature selection method: \textit{f regression}— was employed to evaluate the performance of the stacking model with all the other relevant regressor models. During this process, some features of the dataset were eliminated so that certain physical and chemical traits could be given more importance. The performance of the models was then evaluated. As a result, the top 50 features were selected from the 81 attributes in the dataset.

\begin{table}[!ht]
    \centering
    
    \caption{Dataset Pre-processing}
    \label{tab:dataset_preprocessing}
    \scriptsize
    \begin{tabular}{|p{1.5cm}|p{2cm}|p{1.5cm}|p{2.1cm}|}
        \hline
        \textbf{\nohyphens{Feature Selection}} & \textbf{Scaling Method} & \textbf{Number of Features} & \textbf{\nohyphens{Status of Hyperparameter}} \\
        \hline
        None & Min-Max Scaler & 81 & Without optimization \\
        \hline
        None & Min-Max Scaler & 81 & Optimized \\
        \hline
        f\_regression & Min-Max Scaler & 50 & Without optimization \\
        \hline
        f\_regression & Min-Max Scaler & 50 & Optimized \\
        \hline
    \end{tabular}
\end{table}

\subsection{Stacking Method}
Stacking is a machine learning ensemble algorithm also referred to as "stacked generalization." To determine the best way to combine the predictions from more than one primary machine learning algorithm, it uses a meta-learning algorithm \cite{b24}. For a classification or regression problem, stacking has the benefit of combining the best aspects of several efficient models, which results in predictions that outperform those of any individual model in the ensemble \cite{b17}.
In the stacked model, five algorithms were used as estimators and the Random Forest regressor associated with default hyperparameters served as the final estimator. The hyperparameters for the estimators were tuned for individual algorithms once on all the features and once on the reduced feature set, which can be seen in Tables 2 and 3. The five algorithms used as estimators were Ridge, Lasso, KNN, SVR, and MLP. Fig. 1 also represents the architecture of the stacking ensemble algorithm.

\begin{figure}
    \centering
    \includegraphics[width=\linewidth]{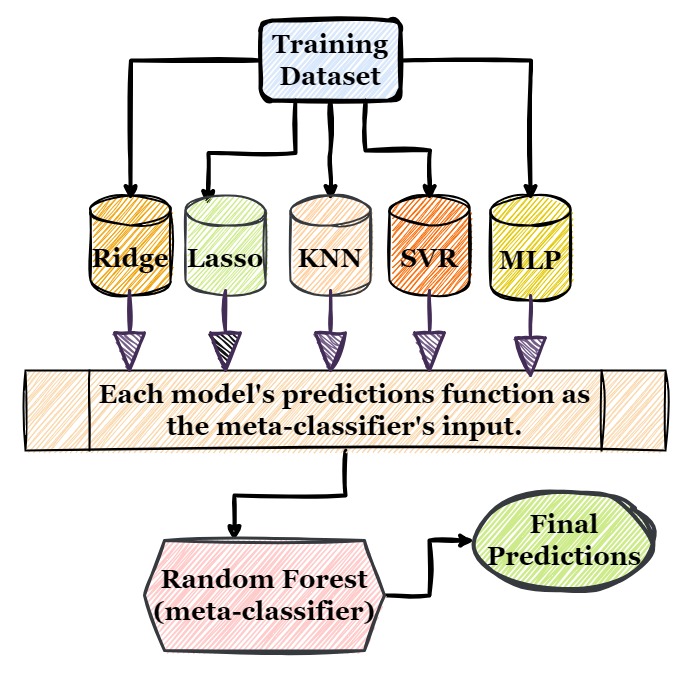}
    \caption{Architecture of Stacking Ensemble method}
    \label{fig:enter-label}
\end{figure}

\subsection{	Hyper-Parameter Optimization}
Hyperparameters in machine learning are all those factors that are deliberately defined by the user to regulate the learning process. These hyperparameters are used to optimize the model's learning process, and their values are specified prior to learning. Therefore, choosing the appropriate set of hyperparameters is essential. Hence, the RandomizedSearchCV was utilized to choose the optimal hyperparameters for this process. There are two methods used by RandomizedSearchCV: "score" and "fit."  The parameters of the estimators are optimized via a cross-validated search across parameters, and these methods are then assessed. The approach here has incorporated hyper parameter optimization for two different cases. Firstly, hyper parameters were optimized for all 81 features. Secondly, only the selected 50 features were set to pass through the hyperparameter optimization process. The best hyperparameters are presented in Table 2 and 3 along
with a list of parameters that were utilized to achieve the best results.

\begin{table*}[htbp]
\centering
\caption{Hyperparameter Optimization On All The 81 Features}
\begin{tabular}{|p{2.5cm}|p{3.5cm}|p{5cm}|p{2.5cm}|}
\hline
\textbf{Algorithm} & \textbf{Hyperparameters} & \textbf{Test Values} & \textbf{Best Vales}\\
\hline
Ridge Regressor & alpha & 0.1, 1, 10, 0.001, 100 & 0.001\\
\cline{2-4}
& tol & 0.001, 0.0001, 0.01, 0.1, 0.00001 & 0.00001\\
\cline{2-4}
& solver & auto, svd, cholesky, lsqr, sparse\_cg, sag, saga & sparse\_cg \\
\cline{2-4}

\hline
Lasso Regressor & alpha & 0.1, 1, 10, 0.001, 100 & 0.001\\
\cline{2-4}
& tol & 0.001, 0.0001, 0.01, 0.1, 0.00001, 0.000001, 0.0000001 & 0.0000001\\
\cline{2-4}

\hline
KNN & n\_neighbors & 2, 5, 10, 25, 50  & 2\\
\cline{2-4}
& leaf\_size & 10, 20, 30, 60, 90, 105, 120, 150 & 20\\
\cline{2-4}
& algorithm & auto, ball\_tree, kd\_tree, brute & brute \\
\cline{2-4}
& p & 1, 2, 3, 5, 10, 20, 40, 80, 100, 200 & 2\\
\cline{2-4}

\hline
SVR & epsilon & 0.01, 0.1, 1, 10, 100 & 1\\
\cline{2-4}
& C & 0.5, 1, 5, 10, 100, 0.05  & 100\\
\cline{2-4}
& cache\_size & 0.2, 2, 20, 200, 2000 & 2000\\
\cline{2-4}
& coef0 & 0.01, 0.1, 0, 1, 10 & 10\\
\cline{2-4}
& degree & 1, 2, 3, 4, 5 & 1\\
\cline{2-4}

\hline
MLP & activation & logistic, relu & relu\\
\cline{2-4}
& learning\_rate\_init & 0.01, 0.1, 0.001 & 0.1\\
\cline{2-4}
& hidden\_layer\_sizes & (55, 52, 78, 30), (56, 32, 25), (57, 40, 52, 75, 60) & (56, 32, 25) \\
\cline{2-4}

\hline
RFR & n\_estimators & 20, 40, 60, 80, 100, 120 & 120\\
\cline{2-4}
& min\_samples\_split & 2, 4, 8, 10 & 2 \\
\cline{2-4}
& max\_depth & 5, 10, 15, 20 & 20\\
\cline{2-4}

\hline
\end{tabular}
\end{table*}

\begin{table*}[htbp]
\centering
\caption{Hyperparameter optimization On 50 Features After Feature Reduction}
\begin{tabular}{|p{2.5cm}|p{3.5cm}|p{5cm}|p{2.5cm}|}
\hline
\textbf{Algorithm} & \textbf{Hyperparameters} & \textbf{Test Values} & \textbf{Best Vales}\\
\hline
Ridge Regressor & alpha & 0.1, 1, 10, 0.001, 100  & 0.001\\
\cline{2-4}
& tol & 0.001, 0.0001, 0.01, 0.1, 0.00001 & 0.1\\
\cline{2-4}
& solver & auto, svd, cholesky, lsqr, sparse\_cg, sag, saga & svd\\
\hline
Lasso Regressor & alpha & 0.1, 1, 10, 0.001, 100 & 0.001\\
\cline{2-4}
& tol & 0.001, 0.0001, 0.01, 0.1, 0.00001, 0.000001, 0.0000001 &0.000001\\
\hline
KNN & n\_neighbors & 2, 5, 10, 25, 50 & 2 \\
\cline{2-4}
& leaf\_size & 10, 20, 30, 60, 90, 105, 120, 150 & 20 \\
\cline{2-4}
& algorithm & auto, ball\_tree, kd\_tree, brute & ball\_tree\\
\cline{2-4}
& p & 1, 2, 3, 5, 10, 20, 40, 80, 100, 200 & 1 \\
\hline
SVR & epsilon & 0.01, 0.1, 1, 10, 100 & 1 \\
\cline{2-4}
& C & 0.5, 1, 5, 10, 100, 0.05 & 100\\
\cline{2-4}
& cache\_size & 0.2, 2, 20, 200, 2000 & 20 \\
\cline{2-4}
& coef0 & 0.01, 0.1, 0, 1, 10 & 0.1\\
\cline{2-4}
& degree & 1, 2, 3, 4, 5 & 1\\
\hline
MLP & activation & logistic, relu & relu\\
\cline{2-4}
& learning\_rate\_init & 0.01, 0.1, 0.001 & 0.001\\
\cline{2-4}
& hidden\_layer\_sizes & (55, 52, 78, 30), (56, 32, 25), (57, 40, 52, 75, 60) & (55, 52, 78, 30)\\
\hline
RFR & n\_estimators & 20, 40, 60, 80, 100, 120 & 120\\
\cline{2-4}
& min\_samples\_split & 2, 4, 8, 10 & 4\\
\cline{2-4}
& max\_depth & 5, 10, 15, 20 & 20\\
\hline
\end{tabular}
\end{table*}

\subsection{Workflow}

The procedure of pre-processing the data before applying machine learning algorithms is vital. In this work, the performance of the \textit{f regression} feature selection technique was compared with the baseline. First, the dataset was scaled using the \textit{min max} scaler from sklearn. In order to find the base results, first all 81 features were used. Furthermore, five standalone machine learning techniques were utilized to estimate the critical temperature: Ridge, Lasso, KNN, SVR, and MLP. Additionally, two separate models, the stacking ensemble model and the voting ensemble model, were utilized to estimate the critical temperature. All the models were first run with the default hyperparameters. Later, RandomizedSearchCV algorithm was used to optimize the hyperparameters of the five standalone models. The estimators in the stacking and voting classifiers used the same hyperparameters. After training the machine learning models, the results were tested using two evaluation metrics, the root mean squared error (RMSE) and R2 score, which are often used in statistical analysis and machine learning to assess the effectiveness of the models. Both the results with the default hyperparameters and the optimized hyperparameters are shown in Table 4 and 5.
In the subsequent phase of this work, the \textit{f regression} feature selection technique was utilized. Using the \textit{min max} scaler from sklearn library, the data was initially normalized. After executing the \textit{f regression} method, a new dataset with the top 50 features was generated. On this new feature set, the same machine learning models as in the baseline were applied. Similarly, this time all models were executed with the default and optimized hyperparameters for 81 features. Fig. 2 shows the comprehensive workflow diagram.

\begin{figure}
    \centering
    \includegraphics[width=0.95\linewidth]{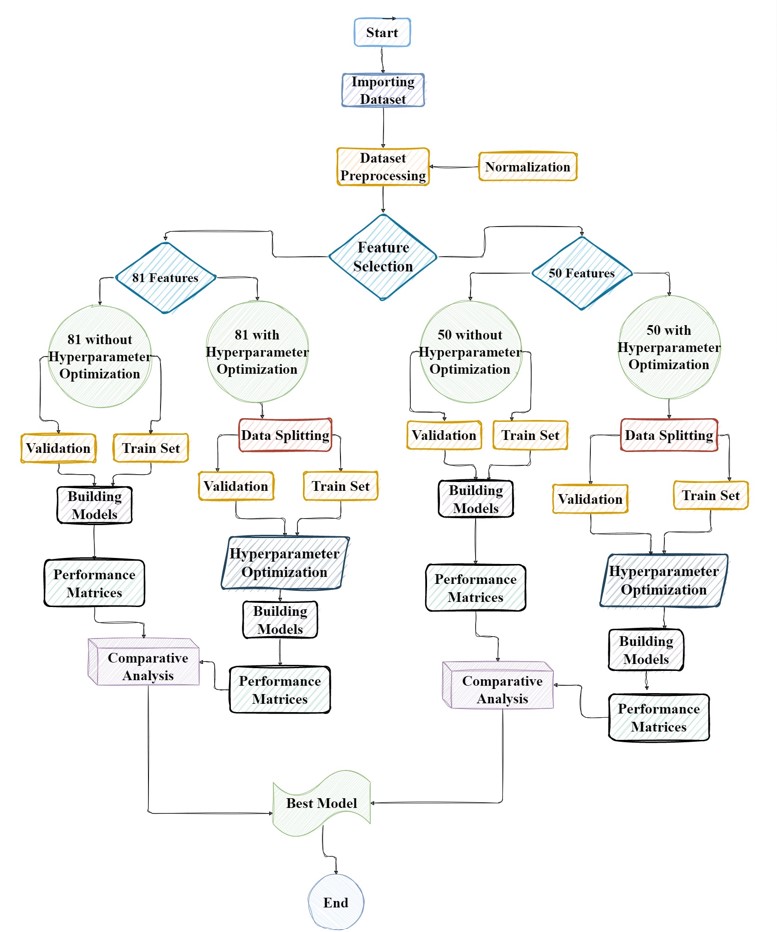}
    \caption{Overall workflow diagram.}
    \label{fig:enter-label}
\end{figure}

\vspace{2\baselineskip}

\section{Results}
\subsection{RMSE Average Result }
The stacking model performed better than other regressor models in terms of RMSE and R2 scores. Table 4 shows the average RMSE values under four separate conditions, in which the stacking model shows the best result of RMSE 9.68684 after hyperparameter optimization with all 81 features taken into consideration. Interestingly, the stacking model indicates a much more consistent result whereas all the other relevant ML models show variations at different conditions as shown in Fig. 3. 
\begin{figure}[h]
    \centering
    \includegraphics[width=\linewidth]{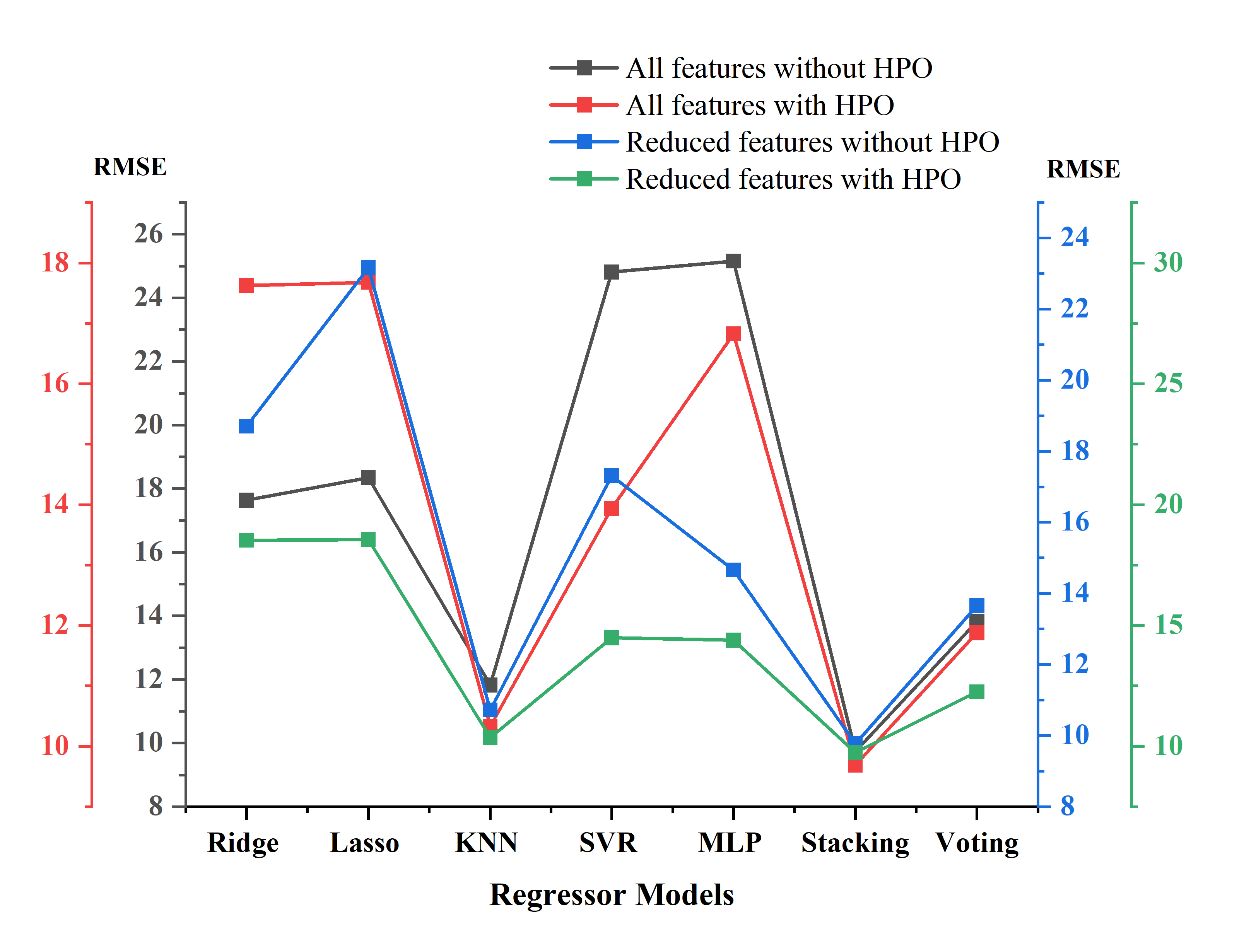}
    \caption{Performance comparison of all the models for RMSE}
    \label{fig:enter-label}
\end{figure}

\subsection{R2 Average Result  }
The stacking model performed very well in the context of R2 scores where the best R2 score was 0.91996 among the four different approaches that were tested in this study. Table 5 shows the relevant comparison of the model with all other ML algorithms taken into consideration. In Fig. 4 the consistency of the stacking model with respect to other regressor models can be observed once again.

\begin{figure}
    \centering
    \includegraphics[width=\linewidth]{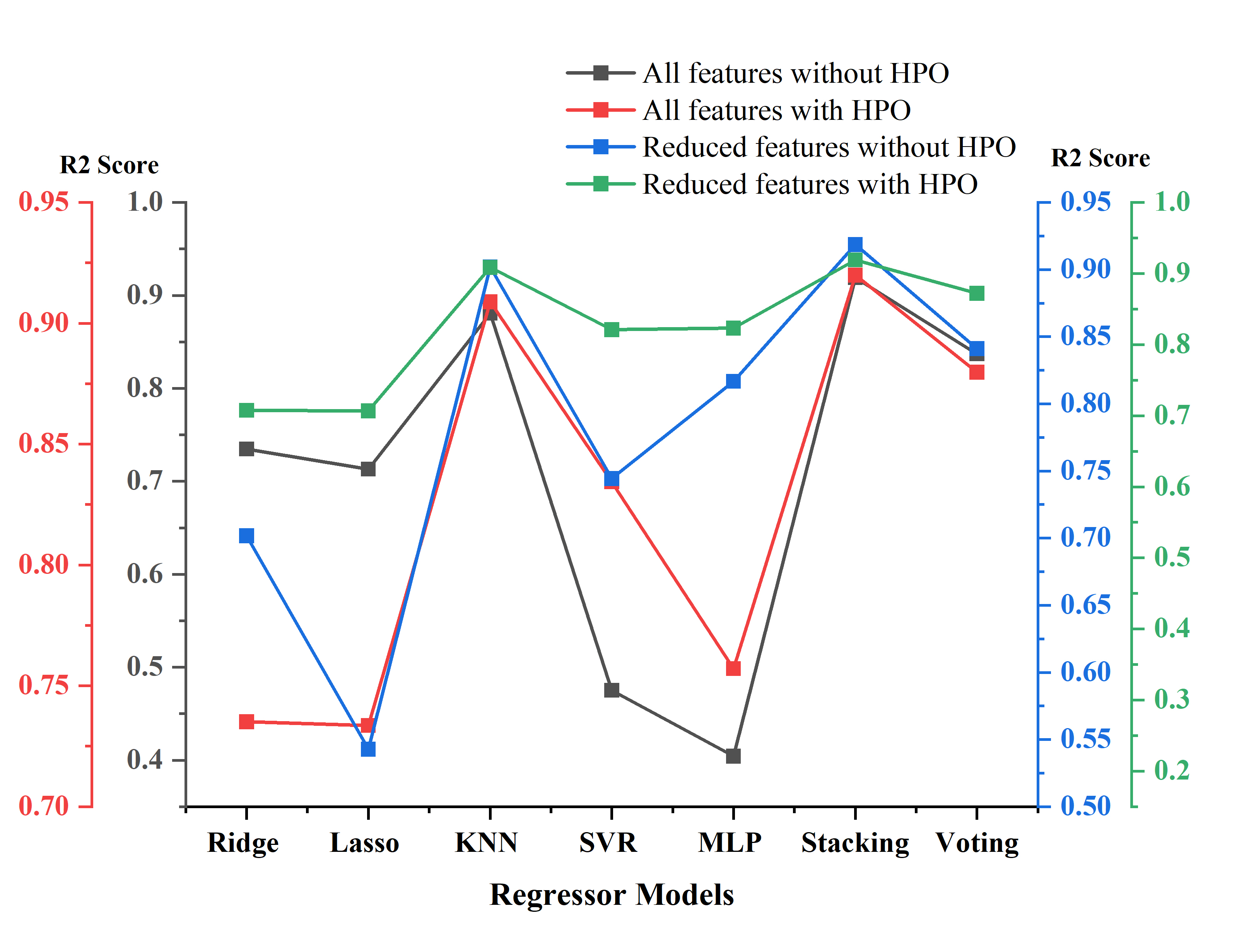}
    \caption{Performance comparison of all the models for R2 score}
    \label{fig:enter-label}
\end{figure}

\begin{table}[!ht]
    \scriptsize
    
    \centering
    \caption{ Average RMSE Under All Conditions}
    \begin{tabular}{|p{1.10cm}|p{1.5cm}|p{1.25cm}|p{1.6cm}|p{1.25cm}|}
    \hline
    \begin{tabular}{@{}c@{}}
    \textbf { ML } \\
    \textbf { Models }
    \end{tabular} &
    \begin{tabular}{@{}c@{}}
    \textbf { All Features } \\
    \textbf { Without } \\
    \textbf { HPO }
    \end{tabular} &
    \begin{tabular}{@{}c@{}}
    \textbf { All } \\
    \textbf { Features } \\
    \textbf { With HPO }
    \end{tabular} &
    \begin{tabular}{@{}c@{}}
    \textbf { Reduced } \\
    \textbf { Features } \\
    \textbf { Without HPO }
    \end{tabular} &
    \begin{tabular}{@{}c@{}}
    \textbf { Reduced } \\
    \textbf { Features } \\
    \textbf { With HPO }
    \end{tabular} \\
    \hline
    \text { Ridge } & 17.6367 & 17.6277 & 18.7059 & 18.5215 \\
    \hline
    \text { Lasso } & 18.3454 & 17.6778 & 23.1596 & 18.5481 \\
    \hline
    \text { KNN } & 11.8217 & 10.3327 & 10.7186 & 10.356 \\
    \hline
    \text { SVR } & 24.8124 & 13.9382 & 17.3152 & 14.4879 \\
    \hline
    \text { MIP } & 25.1586 & 16.8264 & 14.6578 & 14.388 \\
    \hline
    \textbf { Stacking } & \textbf{9.73425} & \textbf{9.68684} & \textbf{9.76847} & \textbf{9.73199} \\
    \hline
    \text { Voting } & 13.8267 & 11.8755 & 13.6528 & 12.2521 \\
    \hline
    \end{tabular}
\end{table}

\begin{table}[!ht]
    \centering
    \scriptsize
    \caption{ Average R2 Score Under All Conditions}
    \begin{tabular}{|p{1.10cm}|p{1.5cm}|p{1.25cm}|p{1.6cm}|p{1.25cm}|}
    \hline
    \begin{tabular}{@{}c@{}}
    \textbf { ML } \\
    \textbf { Models }
    \end{tabular} &
    \begin{tabular}{@{}c@{}}
    \textbf { All Features } \\
    \textbf { Without } \\
    \textbf { HPO }
    \end{tabular} &
    \begin{tabular}{@{}c@{}}
    \textbf { All } \\
    \textbf { Features } \\
    \textbf { With HPO }
    \end{tabular} &
    \begin{tabular}{@{}c@{}}
    \textbf { Reduced } \\
    \textbf { Features } \\
    \textbf { Without HPO }
    \end{tabular} &
    \begin{tabular}{@{}c@{}}
    \textbf { Reduced } \\
    \textbf { Features } \\
    \textbf { With HPO }
    \end{tabular} \\
    \hline
    \text { Ridge } & 0.73487 & 0.73514 & 0.70175 & 0.70759 \\
    \hline
    \text { Lasso } & 0.71313 & 0.73363 & 0.54281 & 0.70675 \\
    \hline
    \text { KNN } & 0.88084 & 0.90892 & 0.90201 & 0.90852 \\
    \hline
    \text { SVR } & 0.47518 & 0.83437 & 0.74443 & 0.82104 \\
    \hline
    \text { MLP } & 0.40419 & 0.75716 & 0.81677 & 0.82318 \\
    \hline
    \textbf{{ Stacking }} & \textbf{0.91915} & \textbf{0.91996} & \textbf{0.91863} & \textbf{0.91922 }\\
    \hline
    \text { Voting } & 0.83704 & 0.87978 & 0.84112 & 0.87203 \\
    \hline
    \end{tabular}
\end{table}

\section{Discussion}
The RMSE values for several algorithms in four different scenarios that forecast the critical temperature are shown in Table 4. The best performance model, the stacking model, attained an RMSE of 9.686 with 81 features and optimized hyperparameters. Feature selection using \textit{f regression} resulted in fewer features, but the stacking algorithm still outperformed others with an RMSE of 9.731. Lasso and Ridge regressor performed lowest, with RMSE scores of 18.548 and 18.521, respectively. Table 5 displays R2 scores for algorithms with 81 and 50 features, respectively, with default and optimized hyperparameters. The stacking method consistently had the greatest R2 score, reaching 0.9199 with 81 features using HPO. Utilizing 81 features and HPO, KNN was the second-best performer achieving an R2 score of 0.9089. The R2 scores for all algorithms improved with optimized hyperparameters for both feature sets. \\In terms of performance metrics, Table 6 displays a comparison of the proposed method with related literature.

\begin{table}
    \centering
    \scriptsize
    \caption{ Comparison Of The Performance Metrics}
    \begin{tabular}{|p{3.5cm}|p{2cm}|p{2cm}|}
    \hline
    \textbf { Research Works } & \textbf { RMSE } & \textbf { R2 score } \\
    \hline
    Li et al. \cite{b18} & 83.565 & 0.899 \\
    \hline
    García-Nieto et al. \cite{b9} & 15.14 & 0.8005 \\
    \hline
    Hamidieh et al. \cite{b15} & 9.5 & 0.920 \\
    \hline
    Babu et al. \cite{b19} & 17.68 & 0.7396 \\
    \hline
    Agbemade. \cite{b11} & 12.01  & 0.8823  \\
    \hline
    Moscato et al. \cite{b20} & 10.989 & --- \\
    \hline
    \textbf{This Work} & \textbf{9.68} & \textbf{0.922} \\
    \hline
    \end{tabular}
\end{table}

\vspace{1\baselineskip}
\section{Conclusion}
This study highlights the potential of ML algorithms for predicting the critical temperature of superconductors based on those elements' chemical and physical make-ups. Some Tc prediction methods have been implemented in this study that includes feature selection, hyperparameter optimization, and integrated learning techniques. The models created for this study exhibit significant prediction capability and may provide new insights into the principles underlying superconductivity in various groups of materials. Researchers interested in discovering room-temperature superconductors might apply the model to constrain their search. However, one shortcoming of this work is that the hyperparameters were found using only RandomizedSearchCV. There might be more effective hyperparameter combinations since this only evaluates a small number of possible combinations. Incorporating physical predictors such as crystal structure or electrical characteristics may enhance and improve the accuracy of our proposed stacking model.

\vspace{1\baselineskip}
\section*{Acknowledgement}

This research was conducted as part of the undergraduate thesis at the Islamic University of Technology (IUT).
The authors would like to thank Mr. Ahnaf Akif Rahman, a researcher from the University of Windsor, Canada, for sharing his knowledge and expertise on Machine Learning and its applications in order to successfully conduct this research.

\vspace{1\baselineskip}
\section*{Data Availability}
The Superconductivity dataset from the UCI Machine Learning Repository was used in order to support this study and is available at \url{https://archive.ics.uci.edu/ml/datasets/Superconductivty+Data}. The prior study and dataset are cited at relevant places within the text as Ref [20].
\vspace{1\baselineskip}
\section*{Conflict of Interest}

The authors do not declare any potential conflict of interest that may alter the outcomes of this
study in any manner.

\vspace{1\baselineskip}

\section*{Copyright}
© [2023] IEEE. Permission from IEEE must be obtained for all other uses, in any current or future research, creating new collective works,  or reuse of any copyrighted component of this work in other works.

\vspace{4\baselineskip}

\end{document}